\documentclass[prl,twocolumn,superscriptaddress]{revtex4}

\usepackage[dvips]{graphicx}%
\usepackage{bm,color}
\usepackage{amsmath,amssymb}

\newcommand{\ketbra}[2]{\ket{#1}\bra{#2}} 
\newcommand{\ket}[1]{\left |  #1 \right \rangle}
\newcommand{\bra}[1]{ \left \langle #1  \right |}
\newcommand{\ave}[1]{ \left\langle #1  \right\rangle}
\def \tr{{\textrm {Tr}}}

\begin{document}
\title{Quantum Benchmark via an Uncertainty Product of  Canonical Variables}

\author{Ryo Namiki}\email[Electric address: ]{namiki@scphys.kyoto-u.ac.jp}%
\affiliation{Department of Physics, Graduate School of Science, Kyoto University, Kyoto 606-8502, Japan}
\affiliation{Institute for Quantum Computing  and Department of Physics and Astronomy,
University of Waterloo, Waterloo, Ontario, N2L 3G1, Canada}
\author{Koji Azuma}\affiliation{NTT Basic Research Laboratories, NTT Corporation, 3-1 Morinosato Wakamiya, Atsugi, Kanagawa 243-0198, Japan}

\date{\today}
\begin{abstract} 
We present an uncertainty-relation-type quantum benchmark for continuous-variable (CV) quantum channels that works with an input ensemble of Gaussian distributed coherent states and homodyne measurements. It determines an optimal trade-off relation between canonical quadrature noises unbeatable by entanglement breaking  
 channels and refines the notion of two quantum duties introduced in the original papers of CV quantum teleportation. It can verify the quantum-domain performance for all one-mode Gaussian channels. 
  We also address the case of stochastic channels and the effect of asymmetric gains.   
\end{abstract}

\maketitle



The quantum benchmarks (QB) \cite{Pop94,Mass95,Horo99,Owari08,Has10,namiki07,Ham05,Cal09} provide a fundamental criterion for  experimental success of quantum gates and channels by  eliminating the possibility that  the physical process is described by an entanglement breaking (EB) channel \cite{16}. This criterion ensures that the channel is capable of transmitting  quantum entanglement and draws a firm bottom line on implementation of quantum gates based on the notion of entanglement.
 The QBs enable us to demonstrate an advantage of entanglement in quantum teleportation process and a non-classical performance in quantum memories  \cite{Ham05,Jul04,Furusawa98,Namiki12R}. 
 They also give a  prerequisite for quantum key distribution \cite{Rig06,namiki04,Khan,Kil12}. A common framework is to observe a response of the gate operation  for a set of non-orthogonal input states.  Currently, the majority of QBs have been given in terms of an average fidelity \cite{Fuc03,namiki08,Namiki12R} and a main theoretical task is to determine the classical limit of the fidelity achieved by EB channels.  By  surpassing  such a classical limit,   one can verify  that the channel is in the {\textit{quantum domain}, namely, 　not an EB channel. 

 Although the fidelity is a central tool to certify  the performance of quantum gates in quantum information science, there has been a  general  interest to  invoke  the canonical uncertainty relation or quadrature noises in evaluating continuous-variable (CV) quantum channels  \cite{Furusawa98,Jul04,Gro03b,Bowen03,Bowen03-,namiki04,Het08,Bra00}.
 In such approaches,  an incoherence of  the gate operation can be intuitively explained by the amount of excess quadrature noises above the shot noise limit assuming the transmission of coherent states or minimum uncertainty states.  
Measurements of  canonical  quadratures are also favorable in many of experiments in quantum information with light and atoms  \cite{Bowen03,Bowen03-,RMP82,CV-RMP,Takano10}. 
 In addition, it recalls a primary question in quantum physics whether  a simple trade-off relation holds between the pair of noises provided that  two of non-commuting observables are measured \cite{Stenholm92,leonhardt}.
 However, 
  such an insightful aspect has little been addressed in  quantum benchmarking. 

One can find an outstanding puzzle on the property of quadrature noises induced by  EB maps.  The original papers of CV quantum teleportation  \cite{Furusawa98} suggested that, to validate an entanglement assistance,  the amount of excess noises  has to be smaller than two units of the shot noise, referred to as \textit{two quantum duties} (two quduties).  There is a famous theorem that a single shot-noise unit of excess noise is unavoidable in the simultaneous measurement of canonical quadratures \cite{footnote,Stenholm92,leonhardt}.  By associating this theorem with  another  theorem \cite{16} that any EB channel can be described as a measurement and a following state preparation,  it is fascinating to interpret each of the measurement and the state preparation  processes as  being responsible for  a single shot-noise penalty \cite{fn1}.
However, this interpretation is inaccurate because the penalty  of two quduties has not been proven to be a classical  limit
 unbeatable by {\textit any} EB channel until now. 
 To this end,   the notion of  two quduties is missing  rigorous links to QBs although it sounds highly interesting  \cite{namiki07-}.

On the contrary, a practical CV QB 
 has been established by determining the classical limit fidelity for an input ensemble using coherent states with a Gaussian prior \cite{Bra00,Ham05,namiki07}. This input ensemble called the \textit{Gaussian distributed coherent states}   gives a modest experimental setting to observe an effectively linear gate response in the CV space where the Gaussian prior suppresses  the contribution  of unfeasibly high-energy input states. Although it is customary to describe a normal CV channel with a gain and excess noise terms of canonical quadrature operators \cite{RMP82,CV-RMP}, the scope of the CV QB had  initially been limited for unit-gain channels corresponding to   unitary action of CV channels \cite{Ham05}. It was then extended for non-unit-gain channels in order to deal with an important class of non-unitary processes such as lossy channels and amplification channels  \cite{namiki07}. This extension is sufficient to detect all one-mode Gaussian channels in the quantum domain
   \cite{namiki07}  similarly to the case of the famous sum criterion for CV entanglement that witnesses   all two-mode Gaussian entanglement \cite{Duan,Simon}. Recently,  another extension has been made to serve for probabilistic operations \cite{Chiri13}.
 Hence,  the fidelity-based QB provides a standard method for  estimating CV quantum operations as well as channels. On the other hand, proving a CV process with  the  ensemble of  Gaussian distributed coherent states could be  a more general idea applicable to varieties of measurement scenarios other than the fidelity. However, it is an open question whether such a setting finds a significant utility besides the fidelity-based method. 

In this Letter, we present an EB limit in a product form  of canonical noises  averaged over  Gaussian distributed coherent states.   It offers a QB that generally explains an optimal trade-off relation between the canonical variables rendered by EB maps and refines the notion of  two quduties. Our QB is shown to be sufficient for detecting all one-mode Gaussian channels in 
the quantum domain. 
 We also
 generalize our QB to deal with  probabilistic operations and an asymmetry of quadrature gains. 
Our results almost repeat the fidelity-based achievements  but demonstrate a   fundamental role of canonical variables to observe genuine quantum coherence in a physical process.

Our goal is to derive a bound from  the first and second moments of canonical variables for output states of a given channel ${\cal E}$ by assuming input of coherent states.
We start with the product separable condition \cite{Gio03} in a normalized form \cite{Namiki12J}: 
Any separable state  $J_{AB}$ satisfies 
\begin{eqnarray}
 &&\tr [ (u \hat x_A-  v \hat  x_B )^2 J ] \tr [ (u \hat  p_A+   v \hat   p_B )^2 J ]  \nonumber \\ 
  & \ge  & \ave{\Delta ^2 ( u  \hat x_A-  v  \hat x_B )}_J \ave{\Delta ^2 (u  \hat p_A+   v  \hat p_B )}_J  \ge \frac{1}{4},  \label{prodC} \end{eqnarray} where $(u,v)$ is a real vector with 
  $u^2+ v ^2 =1$   
 and  the canonical variables satisfy $ [\hat x_A, \hat p_A ]  =  [\hat x_B, \hat p_B ]  =i$.   The first inequality is due to the property of variances, $ \ave{ \hat o^2} \ge \ave{\Delta ^2 \hat o}$.

    Let us write $\hat x_B = (\hat b+ \hat  b^\dagger) / \sqrt 2   $ and  $\hat p_B = (\hat b- \hat b^\dagger) /(\sqrt 2 i)  $.
From the cyclic property of the trace we have \begin{eqnarray}
\begin{split} \tr_B [  \hat x_B  J ]  &=   \tr_B [   \hat b ^\dagger  { J} +  {J} \hat b ]/\sqrt 2, & \\
 \tr_B [  \hat x_B^2   J ]   & =  \tr_B [ \hat b ^{\dagger 2 }  {  J} +  { J} \hat b ^2  +2 \hat b ^{\dagger  }  {  J} \hat b -J  ) ]/2  ,& \\
 \tr_B [  \hat p_B  J ]  &=   i \tr_B [   \hat b ^\dagger  { J} -  {J} \hat b ]/\sqrt 2,  &\\
 \tr_B [  \hat p_B^2   J ]   & =  - \tr_B [ \hat b ^{\dagger 2 }  {  J} +  { J} \hat b ^2  - 2 \hat b ^{\dagger  }  {  J} \hat b + J  ]/2.    & \label{nandato}
\end{split}\end{eqnarray} Here,  $\tr _{A (B)}$ denotes the partial trace over subsystem $ A (B)$.    Let us write the partial trace as 
 $\tr_B [\ \cdot \ ] \to \int   \bra{\alpha ^ *} \cdot   \ket{\alpha ^ *}_B  {d^2 \alpha  }/{\pi} $ by the completeness relation for coherent states. Then, the property of the coherent state,
   $ \hat b \ket{\alpha ^ *}_B = \alpha ^*  \ket{\alpha ^ *}_B$, 
    enables us to show   \begin{eqnarray}
 \begin{split} \tr_B  [  \hat x_B  J ]   &=    \int    x_\alpha   \bra{\alpha ^ *} J  \ket{\alpha ^ *}_B   \frac{d^2 \alpha  }{  \pi},&  \\ 
  \tr_B [  \hat x_B^2   J ]  
 &=    \int     x_\alpha ^2    \bra{\alpha ^ *} J  \ket{\alpha ^ *}_B   \frac{d^2 \alpha  }{  \pi} - \frac{J_A}{2}, &  \\ 
 \tr_B  [  \hat p_B  J ]   &=  -  \int    p_\alpha   \bra{\alpha ^ *} J  \ket{\alpha ^ *}_B   \frac{d^2 \alpha  }{  \pi},&  \\ 
  \tr_B [  \hat p_B^2   J ]  
 &=     \int     p_\alpha ^2    \bra{\alpha ^ *} J  \ket{\alpha ^ *}_B   \frac{d^2 \alpha  }{  \pi} - \frac{J_A}{2},&  \end{split} \label{eeee} \end{eqnarray} where $J_A = \tr _B [J ]$ and  we use a shorthand notation of the mean quadratures of a coherent state  as \begin{eqnarray}
x_\alpha  :=  \bra{\alpha}\hat x  \ket{\alpha}= \frac{\alpha + \alpha ^*}{\sqrt 2 }, \  p_\alpha  :=   \bra{\alpha}\hat p \ket{\alpha}= \frac{\alpha - \alpha ^*}{\sqrt 2 i }.  \label{Eq2}\end{eqnarray} 
By substituting Eqs.~(\ref{eeee})  into the first line of Eq.~(\ref{prodC})  we obtain the following Lemma. 

    \textbf{Lemma.---}Any separable state $J_{AB}$ has to satisfy 
\begin{align}&\prod_{z \in \{x, p\}}\left [  \tr_A  \int ( u  \hat z_A-  v  z_\alpha )^2  \bra{\alpha ^* } J \ket{\alpha ^* }_B  \frac{d^2 \alpha}{\pi} -  \frac{v^2}{2} \right] \ge \frac{1}{4}. &\label{th1} \end{align}
 We may call this Lemma the  \textit{hybrid separable condition} because it can be seen as an entanglement detection scheme where homodyne and heterodyne measurements are respectively performed on subsystems $A$ and $B$    \cite{D99}.  Our main result is the following Proposition.

 \textbf{Proposition.---}For any $\lambda > 0 $ and 
 $\eta  > 0$,  any EB channel $\mathcal E $ satisfies the uncertainty limit (See  Fig. \ref{fig1:fig1.eps})  
\begin{eqnarray}
 \left[  \bar V_x  - \frac{\eta }{2 (1+ \lambda )} \right]   \left[  \bar V_p - \frac{\eta }{2 (1+ \lambda )} \right]  \ge \frac{1}{4}\left( 1+ \frac{ \eta }{1+ \lambda }   \right)^2   \label{prop1}, 
\end{eqnarray} where the mean square deviation is defined through 
\begin{eqnarray} \bar V_z   =  \bar V_z (\eta, \lambda)   :=  \tr \int  p_\lambda (\alpha )   (\hat 
z-  \sqrt \eta  z_\alpha )^2   \mathcal E ( \rho _\alpha ) d^2 \alpha,   \label{MSD}    \end{eqnarray} with $\rho_\alpha := \ketbra{\alpha}{\alpha}$
 and the prior Gaussian distribution 
   \begin{eqnarray}
p_\lambda( \alpha ) :=  \frac{\lambda }{\pi} \exp (- \lambda |\alpha |^2 ).  \label{eq2}\end{eqnarray}   
 This prior enables us to neglect the contribution of high energy states with $|\alpha |^2 \gg \lambda ^{-1}$ and  represents a flat distribution  in the limit $ \lambda \to 0$. As we will see, the \textit{gain factor} $\eta$  and the input ensemble of Gaussian distributed coherent states $\{p_\lambda (\alpha) , \rho_\alpha \}_{\alpha \in \mathbb{C}}$ are  naturally introduced from a simple entanglement detection scenario which uses our Lemma on a two-mode state given by applying a quantum channel $\cal E$ to a two-mode squeezed state.  Moreover,  
  the pair   $(\bar V_x, \bar V_p)$ essentially comes from the quadrature correlations in Eq.~\eqref{prodC} and represents the noise terms of $\mathcal E$.   It  can be directly measured by homodyne detection on the output state $\mathcal E (\rho_\alpha)$.  Note that 
    Eq.~(\ref{prop1}) corresponds to the canonical uncertainty relation when $\eta =0$.

\begin{figure}[tbph] \includegraphics[width=0.65\linewidth]{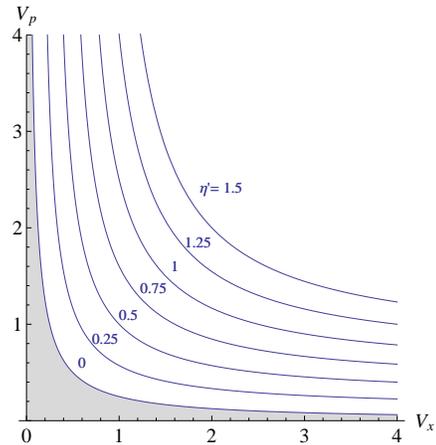} 
 \caption{ Entanglement breaking limit for quadrature noises   in Eq.~(\ref{prop1}) for  a set of a normalized gain $\eta' = \eta / (1+ \lambda)$.  The case of $\eta =0 $ retrieves the minimum uncertainty curve, and the shade represents physically unaccessible regime.}  \label{fig1:fig1.eps}\end{figure}

 \textit{Proof of Proposition.---}
Suppose that the bipartite state $J$ is prepared by the action of a channel $\mathcal E$  as  \cite{Namiki11a,Namiki11R}
\begin{eqnarray}
 J&=&  \mathcal E_A \otimes I_B \left( \ket{\psi_\xi}\bra{\psi_\xi} \right ), \label{CJstate}\end{eqnarray}
where  $\ket{\psi_\xi}= \sqrt{1-\xi ^2} \sum_{n=0}^\infty \xi^n\ket{n}\ket{n}$ with   $\xi \in (0,1)$  is a two-mode squeezed state and  $I$ represents the identity process. This implies 
 \begin{eqnarray}
 \bra{\alpha ^*}J \ket{\alpha ^* }_B  = (1-\xi^2) e ^{-(1-\xi^2)|\alpha |^2 }\mathcal E_A( \rho_{ \xi\alpha}  ) . \label{teq}\end{eqnarray}
From Eqs.~(\ref{eeee}),   (\ref{MSD}), (\ref{eq2}), and (\ref{teq}) we can write 
\begin{eqnarray}  & &\tr_A  \int ( u  \hat z_A-  v  z_\alpha )^2   \bra{\alpha ^* } J \ket{\alpha ^* }_B  \frac{d^2 \alpha}{\pi}  \nonumber \\&= &  u^2  \left[ \tr_A  \int p_{\lambda}(\alpha ) (  \hat z_A-  \sqrt \eta  z_\alpha )^2 \mathcal E_A( \rho_\alpha )   {d^2 \alpha} \right]
  =  u^2  \bar V_z,  \nonumber \\   \label{slast} \end{eqnarray}
 where $z \in \{x, p\}$, the variable of the integration is rescaled as $\xi \alpha \to  \alpha  $,  and new parameters are given by 
 \begin{eqnarray}
 \lambda = \frac{1- \xi^2}{\xi^2 } >0 , \ \eta = \frac{1}{\xi ^2 }\left( \frac{v}{u} \right) ^2 \ge 0. \label{defLE}
\end{eqnarray}
If we eliminate $\xi$ in Eqs.~(\ref{defLE}), we have 
 \begin{eqnarray}
\frac{1}{u^2} = \left( 1 + \frac{\eta }{1+ \lambda }\right)  , \textrm{ or equivalently  }  \frac{v^2}{u^2 }=  \frac{\eta }{1+ \lambda }. 
\label{defLE2}
\end{eqnarray}
Now, suppose that $\mathcal E$ is an EB channel. Then $J$ of Eq.~(\ref{CJstate}) is a separable state, and we can use Lemma. By substituting Eq.~(\ref{slast}) into Eq.~(\ref{th1}) we have
\begin{eqnarray}
 u^4  \left (\bar V_x   - \frac{(v/u)^2}{2} \right)  \left (\bar V_p   -  \frac{(v/u)^2}{2} \right) \ge \frac{1}{4}. 
\end{eqnarray} Finally, substituting Eq.~(\ref{defLE2}) into this expression we obtain Eq.~(\ref{prop1}).  \hfill$\blacksquare$

  Any violation of the condition of  Eq.~(\ref{prop1}) implies that the channel cannot be simulated by EB channels, and it establishes a QB to verify the quantum-domain process with the input ensemble of Gaussian distributed coherent states and normal quadrature measurements. Main implication of our QB is the followings: For unit gain $\eta = 1$ and completely unknown coherent states $\lambda \to 0$, Eq.~(\ref{prop1})  reduces to $(\bar V_x - V_0)(\bar V_p - V_0) \ge (2 V_0)^2$ where  $V_0 := \ave{\Delta^2 \hat z}_{\rho_\alpha}= 1/2$ is the variance of coherent states or the shot noise. This implies $\bar V_z-V_0 \ge 0$ represents the extra noise added by the channel ${\mathcal E}$.  Then, the inequality $(\bar V_x-V_0)(\bar V_p-V_0) \ge (2V_0)^2$ states that the product of the extra noises is not less than $(2V_0)^2$. This $2V_0$ coincides with  two quduties  \cite{Furusawa98} which have been introduced as the extra noise induced by the classical teleportation \cite{Bra00,fn1}.  However, note that our bound  $(\bar V_x-V_0)(\bar V_p-V_0) \ge (2V_0)^2$  reveals a more fundamental aspect of two quduties, that is,  two quduties  $2V_0$ correspond to the minimum of extra noises induced by {\it arbitrary} EB channels. 
  Moreover, the role of the product form is striking.    In general we could observe 
    $\bar V_x \neq  \bar V_p$, 
   and     some of EB maps induce an extra noise for one quadrature, say $\hat x$,   so that it keeps below two quduties as $\bar V_x - V_0 <2 V_0 $. However, even in such cases, our  formula  states that the  extra noise of the other quadrature $\bar V_p -V_0 $ has to increase to fulfill the limit in the product form.  Therefore, the classical penalty on the canonical variables  is demonstrated  as a fundamental basis  through the \textit{uncertainty product} and it refines the notion of two quduties.  This fundamental structure holds for non-unit gain $\eta >0$ and partially known coherent states $\lambda >0$.  A non-unity gain  $\eta \neq 1$ implies the amplitude transformation $\alpha \to \sqrt \eta \alpha$. Thereby, the minimum of the uncertainty product has to keep the scaling determined  by the gain factor $\eta $ similarly to the amplification-uncertainty principle \cite{amp}.  For a finite distribution,  $\lambda^{-1}$ represents the width of the prior $p_\lambda$ of Eq.~(\ref{eq2}).   Hence, the factor $1+ \lambda$ of Eq.~(\ref{prop1}) is thought to be the reduction of the uncertainty due to   the amount of prior knowledge. 

   Interestingly, one can find an EB map that achieves the equality of Eq.~(\ref{prop1}) for any possible parameter set of $(\eta, \lambda)$.  This means that Eq.~(\ref{prop1}) is tight for every pair of $(\eta, \lambda)$  and  the inversely proportional curves of Fig.~\ref{fig1:fig1.eps} entirely  describe  an optimal trade-off relation between canonical quantum noises to beat the classical channels.  In fact, we can show Eq. (\ref{prop1}) is saturated by the EB map of
 \begin{eqnarray} \mathcal E_{MP} (\rho) = \int S_R \ketbra{ \gamma \alpha }{\alpha } S_r^\dagger  \rho   S_r \ketbra{  \alpha }{\gamma \alpha } S_R^\dagger   \frac{d^2 \alpha }{ \pi},  \label{MPmap} \end{eqnarray} where  $ S_r = e^{r(\hat a^2- \hat a^{\dagger 2 })/2}$ is a squeezer and  \begin{align}\begin{split} \gamma  &=  \frac{\sqrt \eta }{\sqrt{(1+\lambda )^2 \cosh^2r - \sinh^2 r }}, \\    e^R  & =  \sqrt\frac{{ (1+ \lambda )\cosh r + \sinh  r }}{ {(1+ \lambda ) \cosh r -\sinh  r }} .\label{gammam} \end{split} \end{align}
This yields a simple form $(\bar V_x ,\bar V_p )= u^{-2}(e^{-2R}+ v^2,e^{2R}+ v^2 )/2$ with  Eq.~(\ref{defLE2}),
 and $R$ determines the balance between $\bar V_x$ and $\bar V_p$. 
Obviously, $\mathcal E_{MP} $ represents the channel that prepares a minimum uncertainty state  after  a projection to a minimum uncertainty state. This structure demonstrates  the mechanism that  each of the measure and  preparation processes is responsible for  increasing the excess noises by  a single quduty. 

 As we will prove next, our QB is enough to detect  all  one-mode Gaussian channels in the quantum domain. This is reasonable because $J$ is a Gaussian state whenever $\cal E$ is a Gaussian channel and  the separable condition of Eq.~\eqref{prodC} is known to be sufficient for detection of any two-mode Gaussian entanglement. In the proof of Proposition, our input ensemble $\{p_\lambda (\alpha), \rho_{\alpha }\}_{\alpha \in \mathbb{C}}$ is determined by the pair of the two-mode squeezed state $\ket{\psi_\xi}$ and the coherent-state basis $ \{\ket{\alpha}\}_{\alpha \in \mathbb{C}}$.  Although one can consider different input ensembles by assigning other entangled states, it remains open  whether any given ensemble can be  related to a meaningful entanglement  detection scenario \cite{fn2}.  
    On the other hand,      the basis $\{\ket{\alpha}\}_{\alpha \in \mathbb{C} }$ is rather regarded as a choice of  the representation that executes the partial trace, but,  enables us to introduce experimentally relevant input states.

We can show the converse statement of our Proposition for  the class of one-mode Gaussian channels: If $\mathcal E$ is Gaussian and  not EB,  there exists  a set of  $(\eta,\lambda)$ and additional Gaussian unitary operators  so that  $\mathcal E$ violates Eq.~(\ref{prop1}).  This can be proven similarly to the case of the fidelity-based benchmark \cite{namiki07}: Thanks to Holevo's classification of Gaussian channels  \cite{Hol08}, it is sufficient to check that the following two types of the channels violate Eq.~(\ref{prop1}).    One is a unit-gain channel ($\eta =1$) which adds one unit of shot noise to one of quadratures, e.g., $(\bar V_x, \bar V_p) = V_0(2, 1)$.   It violates the condition of  Eq.~(\ref{prop1}) for $\lambda <4 $.  The other is an amplification/attenuation channel which transforms the moments of  both quadratures as $z_\alpha \to \sqrt{G  } z_\alpha $ and  $\ave{\hat z^2}_{\rho_\alpha } \to  G z_\alpha ^2  +\tilde n  + (G + | 1- G |) /2 $ where 
    $G \ge 0$ is an actual  gain and $\tilde n \in [0, \min \{1, G \}) $. This implies $\bar V_z = \lambda^{-1}(\sqrt {G } - \sqrt \eta )^2 + \tilde n +(G + | 1- G |) /2$,  
 and  the condition of  Eq.~(\ref{prop1}) is violated if $(\eta, \lambda )= (4 G, 1 )$.

   Note that, if the channel is assumed to be Gaussian, it is covariant under displacement \cite{Owari08,Guta10}. Then, one can determine any channel parameters through covariance matrices based on input of a single coherent state. However, the displacement covariance is not physically justifiable because it implies that the channel maintains a linear response even for any high energy input state. Hence, we are better off using the Gaussian assumption for channels. In our theorem,  the footing of  Gaussian distributed coherent states  bypasses the Gaussian assumption  and gives us a practical platform to explore effectively linear responses \cite{Namiki11R,Chiri13}. Such a framework  would be crucial  in experiments to deal with Gaussian and non-Gaussian ingredients equally well.

While the product form of uncertainties represents a fundamental boundary, an EB bound focusing on the total quadrature noise $\bar V  := \bar V_x +  \bar V_p $ was known in Ref.~\cite{namiki07}.
We can improve this bound as a corollary.

 \textbf{Corollary 1.---}Let us define  the total noise as $\bar V  := \bar V_x +  \bar V_p $ with  Eq.~(\ref{MSD}). 
 For any $\lambda> 0$ and $\eta >  0$, any EB channel $\mathcal E $ satisfies  \begin{eqnarray} \frac{\bar V }{2}  = \frac{\bar V_x   + \bar V_p}{2}  \ge \frac{1}{2}+  \frac { \eta}{ 1+ \lambda }. \label{Coro1}\end{eqnarray}
This can be proven by applying the relation $|a|   + |b| \ge 2 \sqrt{|a b | } $ to Proposition.  The inequality of Eq.~(\ref{Coro1}) is tight as it can also be saturated by  the EB map $\mathcal E_{MP}$ of Eq.~(\ref{MPmap}) with $r=0$.  It improves the QB inequality of Eq.~(10) in Ref. \cite{namiki07} 
(See \cite{B44}).   Corollary 1 can be associated with the famous sum condition for  separability \cite{Duan} whereas 
   Proposition has its origin in the product separable condition  of Eq.~(\ref{th1}).  
 Note that,  from  the total noise $\bar V$, one can obtain a lower bound of  the average fidelity  $\bar F_{\eta,\lambda}:= \int p_\lambda (\alpha) \bra{\sqrt \eta \alpha }   \mathcal E (\rho_{\alpha}) \ket{\sqrt \eta \alpha} d^2 \alpha $ \cite{Bra00,Ham05,namiki07,Namiki11a,Namiki11R,Chiri13} by using the relation $\bar F_{\eta,\lambda} \ge  {(3 - \bar V)/}{ 2} $ introduced in \cite{namiki07}.  This  supports the intuition that a  smaller excess noise implies a higher fidelity, and simply connects the measurement of  $(\bar V_x,\bar V_p)$   to an estimation of  the fidelity.


Finally, we generalize our Proposition to address   the effects of (i)  asymmetric gains
where the first moments are expected to transform  $(x_\alpha, p_\alpha) \to (g_x x_\alpha, g_p p_\alpha)$ \cite{Bowen03,Bowen03-} and (ii) the post-selection  where the channel can be  a trace-decreasing map (stochastic quantum channel)  \cite{Chiri13}. 

 \textbf{Corollary 2.---}For any $\lambda > 0 $ and any gain pair $(g_x, g_p)  > 0$,  any stochastic EB map $\mathcal E$  satisfies 
 \begin{eqnarray}
 \left[   {\tilde V_{x }} - \frac{g_x^2 }{2(1+\lambda) } \right]   \left[  {\tilde V_{p }}  - \frac{g_p^2 }{2 (1+\lambda) } \right]  \ge \frac{1}{4}\left( 1+ \frac{g_x g_p }{1+\lambda}   \right)^2,   
 \label{coll2}
\end{eqnarray}
where $\tilde V_{z} :=  \bar V_{z} (g_z^2, \lambda) / \tr[ \int p_\lambda (\alpha) {\cal E}( \rho_\alpha )d^2 \alpha ] $ with Eq.~(\ref{MSD}). 

To prove Corollary 2, we replace $ J$ in Eq.~\eqref{prodC} with $  (S_q)_B  J  (S_q ^\dagger)_B $.  This  transforms the  quadratures in Lemma  as $ (\hat x_A, \hat p_A)  \to  (\hat x _A e^{q}, \hat p_Ae^{-q})$.   Further, repeating the proof of Proposition starting with $J = \mathcal E_A \otimes I_B  (\ketbra{\psi_{\xi}}{\psi_\xi})/ \tr[\mathcal E_A \otimes I_B  (\ketbra{\psi_{\xi}}{\psi_\xi})] $ instead of Eq.~\eqref{CJstate} we can reach Corollary 2 with the form of the gain pair  $(g_x, g_p) =  (\sqrt \eta e^{-q},  \sqrt \eta e^q) $. 
Since the underlying physics does not change as long as $J$ is normalized, $\cal E$ is not necessary to be a trace-preserving operation.
 Thus, EB channels  are unable to beat our bound even stochastically. Therefore, Corollary 2 constitutes a unified QB that  works  with feasible input-and-measurement settings for a wide class of CV channels by assigning a gain pair $(g_x, g_p)$.  Corollary 2 also describes an optimal trade-off relation due to EB maps  
 since  the inequality of Eq.~(\ref{coll2}) is saturated by the EB channel  $\mathcal E _{MP}'( \rho )=  S_q \mathcal E _{MP} (  \rho  )  S_q^\dagger$  with Eq.~(\ref{gammam}) for any given $(g_x,g_p,\lambda)>0$. Although one can use the fidelity-based QB \cite{namiki07} for asymmetric gains,   it may require a type of squeezed resources such as a measurement of the fidelities to squeezed states \cite{namiki08}.   

In conclusion, we have established an uncertainty-relation-type QB for CV quantum channels. It is usable to verify the quantum-domain performance for a wide class of CV quantum channels by assigning a pair of quadrature gains including stochastic quantum channels.  Our results  generally explain the classical penalty of two quduties and an optimal trade-off relation on canonical variables to beat EB channels. This highlights a structural difference from the fidelity-based QB  \cite{namiki07}. We have also proven the converse statement of our QB for one-mode Gaussian channels. Hence, our framework has no less generality than the framework of the fidelity-based QB.  
 It would be fundamental to address the quantum-amplification limit \cite{amp,Namiki11R,Chiri13} and related cloning limits  in our canonical basis \cite{NamikiAzumaUP2}. 
Although we have concentrated on a single separable condition  of Eq.~\eqref{prodC}, one can use our approach to  translate    
a wide class of separable conditions \cite{SV,Miran,Namiki12a1} into quantum benchmarking conditions \cite{NamikiUp}.

We thank N. L\"utkenhaus for helpful discussions. 
This work was partly supported by GCOE Program ``The Next Generation of Physics, Spun from Universality and Emergence'' from MEXT of Japan and World-Leading
Innovative R\&D on Science and Technology (FIRST).  
KA is in part supported by the
Project UQCC by the National Institute of Information
and Communications Technology (NICT).


\begin{thebibliography}{}
\bibitem{fn1}  
Note that the noise penalty of two  quduties is given by an EB channel implemented by heterodyne measurement and preparation of coherent states \cite{Bra00}. This model of measure-and-prepare schemes, commonly termed as the  classical teleportation, 
 constitutes an optimal EB map for the fidelity-based QBs \cite{Ham05,namiki07} and gives an  optimal amount of noise tolerance for an entanglement transmission through a generic Gaussian channel (See Eq.~(4) of Ref.  \cite{Has10}).
\bibitem{namiki07-} Roughly speaking,  Eq.~(10) of Ref.~\cite{namiki07} shows that a single quduty works as a QB.  
\bibitem{B44} A similar sum condition appears in  \cite{Bowen03-} and coincides with Eq.~\eqref{Coro1} in the limit $\lambda \to 0$. However, the results in Ref.~\cite{Bowen03-} are based on the conditional variance with some other constraints, and it is an open question whether one can find a rigorous link  to our result.
\end{thebibliography}
\end{document}